# Comparison between Performances of Channel estimation Techniques for CP-LTE and ZP-LTE Downlink Systems

Abdelhakim Khlifi[1] and Ridha Bouallegue[2]

[1]National Engineering School of Tunis, Tunisia
`abdelhakim.khlifi@gmail.com`
[2]Sup'Com, Tunisia
`ridha.bouallegue@gmail.com`

**ABSTRACT**

*In this paper, we propose to evaluate the performance of channel estimation techniques for Long Term Evolution (LTE) Downlink systems based on Zero Padding technique (ZP) instead of Cyclic Prefixing (CP). LTE Downlink system is a multiuser system based on a MIMO-OFDMA technology. Usually, in OFDM systems, a guard interval is inserted in order to mitigate both inter-carrier interference (ICI) and inter-symbol interference (ISI). LTE Downlink systems are based on CP-OFDM technique which consists of a copy a last OFDM symbols inserted at the beginning of each transmitted OFDM symbol. Although this technique shows good performances, the CP-LTE system suffers from a power efficiency loss. With the number of present OFDM symbols in LTE Downlink radio frame, the bandwidth loss becomes more important. Instead of CP, we propose to evaluate the performance of ZP-LTE systems in order to avoid the power efficiency .In this paper, we interest to evaluate the performance of channel estimation techniques for the two LTE Downlink systems. Simulations results show that although ZP-LTE systems outperform CP-LTE Downlink systems in terms of power efficiency, the CP-LTE systems show better performance than ZP-LTE systems and especially for high SNR values. MATLAB Monte–Carlo simulations are used to evaluate the performance of LS, LMMSE and Lr-LMMSE estimators in terms of Mean Square Error (MSE) and Bit Error Rate (BER) for 2x2 LTE Downlink systems.*

## 1. INTRODUCTION

The explosive growth in demand for mobile broadband services has leading to a great need of speed data access systems. Many solutions were proposed in order to serve this demand. MIMO (Multiple-Input Multiple-Output) is the most popular research results field of next-generation mobile communication systems [1]. MIMO has the ability to increase the system capacity without extra-bandwidth [1]. Multipath propagation is one of the most challenging problems in radio mobile transmissions. Multipath propagation usually causes selective frequency channels. As a solution to combat the effect of frequency selectivity of channels, multicarrier modulation has been receiving growing interest in recent years due to simplified equalization in the frequency domain [3]. OFDM is a multi-carrier modulation technique which consists of converting a frequency-selective fading channel into parallel flat-fading sub-channels. The association MIMO-OFDM represents a great achievement in wireless field. The propagation over the radio-frequency channel may affect the transmitted signal with inter- symbol (ISI) and inter-channel (ICI) interferences. Therefore, before transmitting OFDM symbol, a guard interval is added at the beginning of each OFDM symbol in order to mitigate ICI and ISI. The inserted guard interval is usually equal to or longer than to the channel [2].

LTE is the next generation networks responding to the high demand for broadband data access. In this paper, we interest to LTE Downlink systems. LTE Downlink system is a MIMO-OFDMA based system. LTE proposes Orthogonal Frequency Division Multiple Access (OFDMA) as downlink multiple access technique providing 100 Mbps (SISO), 172 Mbps (2x2 MIMO) and 326 Mbps (4x4 MIMO).

The performance evaluation of channel estimation techniques for LTE systems was discussed in many articles e.g [10] [11].

LTE standard use CP as a guard interval inserted at the beginning of each transmitted OFDM symbol. In this paper, we propose to study the performance of channel estimation techniques for LTE Downlink systems based on ZP technique. In the rest of the paper, an overview of LTE Downlink physical layer is given in section II. A LTE MIMO-OFDM system model is given in section III. The LS and the LMMSE and low rank approximation of LMMSE channel estimation techniques are discussed in section IV with simulation results for their performances for the two systems given in section V. Conclusion is given in the last section

## 2. LTE DOWNLINK PHYSICAL LAYER

Figure 1 illustrates the structure of the LTE radio frame. The duration of one LTE radio frame is 10 ms. It is composed of 20 slots of 0.5 ms. Each two consecutive time slots are combined as one sub-frame and so LTE radio frame contains 10 sub-frames.

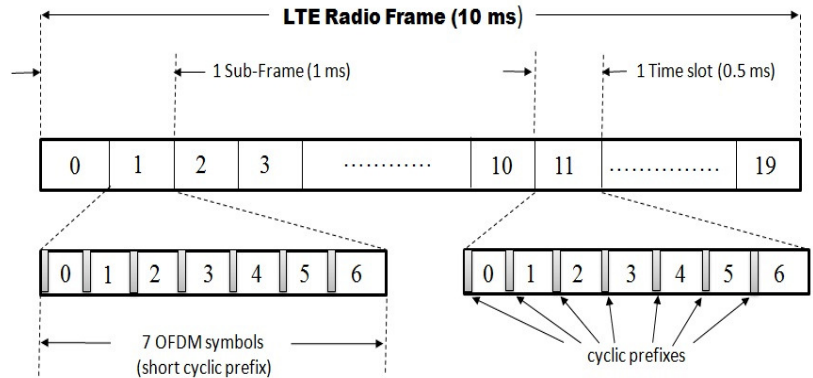

Figure 1: LTE radio Frame structure

The total number of available subcarriers depends on the overall transmission bandwidth of the system as described in table 1. As shown in Figure 2, each PRB occupies a bandwidth of 180 kHz (12 x 15 kHz). Depending on the employed CP; a PRB consists of 84 resource elements (12 subcarriers during 7 OFDM symbols) or 72 resource elements (12 subcarriers during 6 OFDM symbols).

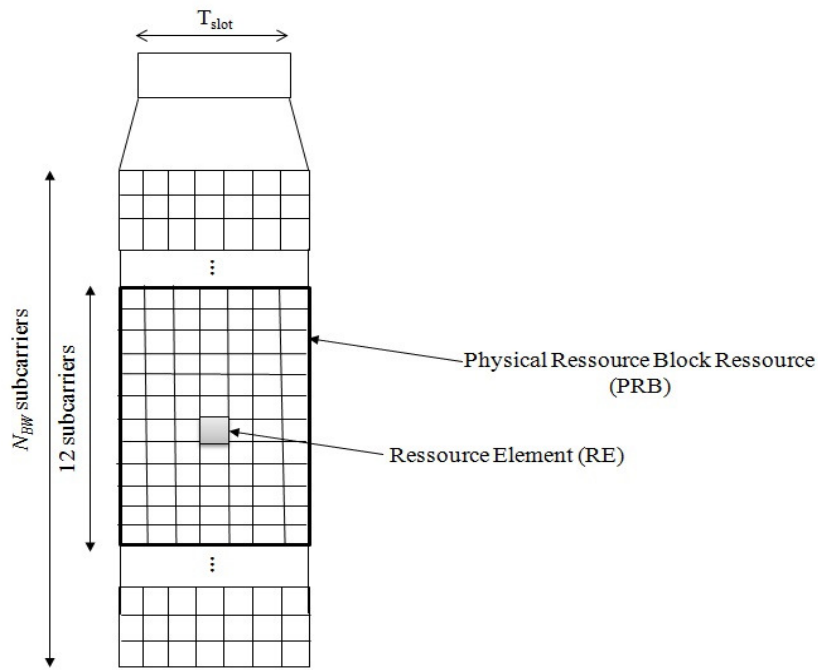

Figure 2: Physical Resource Block

The LTE specifications [5] define parameters for system bandwidths from 1.25 MHz to 20 MHz as shown in Table I. Each bandwidth system has its technical specifications: number of available PRBs, the sampling frequency, FFT size, number of occupied sub-carriers…

TABLE I. LTE DOWNLINK SPECIFICATIONS

| Transmission bandwitdh (MHz) | | 1.25 | 2.5 | 5 | 10 | 15 | 20 |
|---|---|---|---|---|---|---|---|
| Sub-frame duration (ms) | | 0.5 | | | | | |
| Sub-carrier spacing (kHz) | | 15 | | | | | |
| Physical resource block (PRB) | | 180 | | | | | |
| Number of available PRBs | | 6 | 12 | 25 | 50 | 75 | 100 |
| Sampling Frequency (MHz) | | 1.92 | 3.84 | 7.68 | 15.36 | 23.04 | 30.72 |
| FFT size | | 128 | 256 | 512 | 1024 | 1536 | 2048 |
| Virtual Carriers | Left | 28 | 38 | 106 | 212 | 318 | 424 |
| | Right | 28 | 38 | 106 | 212 | 318 | 424 |
| Number of occupied subcarriers | | 72 | 180 | 300 | 600 | 900 | 1200 |

International Journal of Computer Networks & Communications (IJCNC) Vol.4, No.4, July 2012

## 2.1 CP- LTE Downlink system

In this case, and before transmitting each OFDM symbol, a CP is inserted at the beginning of each transmitted OFDM symbol.

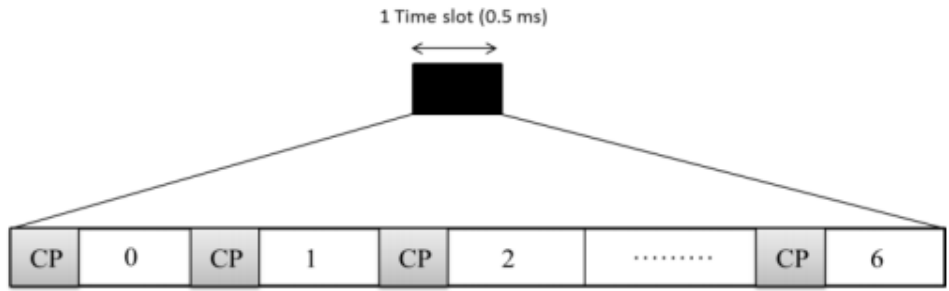

Figure 3: LTE Downlink radio frame based CP

The inserted CP consists of a copy of last part of the OFDM symbol. The last $L_{CP}$ samples of each OFDM symbol of N samples are copied and added in front of the OFDM symbol, as shown in Figure 4.

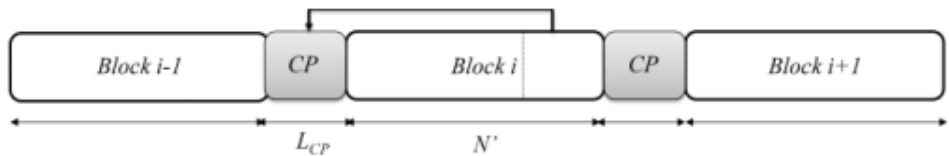

Figure 4:   Transmitted signal for CP-LTE

Because during the guard interval signal is transmitted, the CP-LTE system suffers from a power efficiency loss with a factor of $\frac{N}{N+L_{CP}}$.

## 2.2. ZP- LTE Downlink system

As shown in Fig.5, during this guard interval, no signal is transmitted.

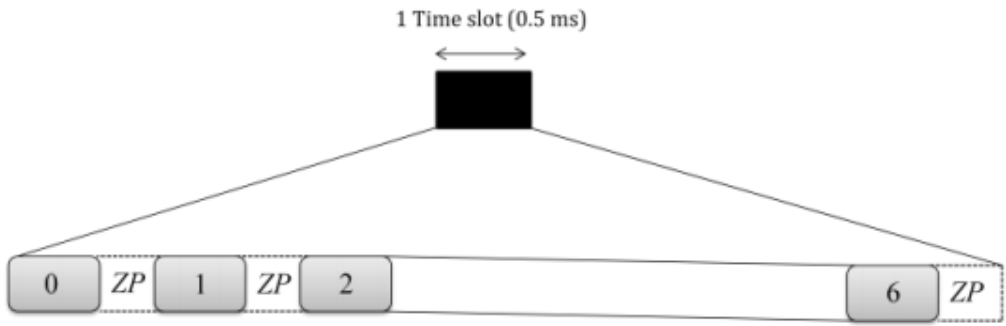

Figure 5:   LTE Downlink radio frame based ZP

The inserted ZP consists of a serie of zeros inserted at the end of each transmitted known with Zero Padding of the OFDM symbol. The last $L_{CP}$ samples of each OFDM symbol of N samples are copied and added in front of the OFDM symbol, as shown in Figure 6.

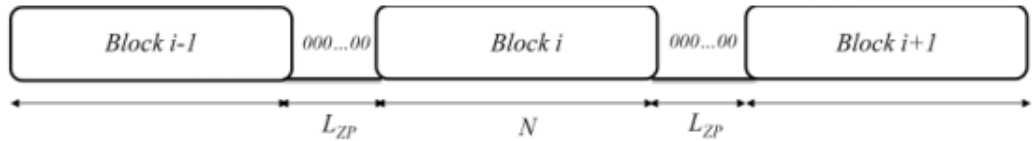

Figure 6: LTE Downlink radio frame based ZP

Although the power efficiency loss is avoided in ZP-OFDM, the noise power will be enhanced with a factor of $\frac{N+L_{ZP}}{N}$.

## 3. LTE DOWNLINK SYSTEM MODEL

The system model is given in Figure 7. LTE Downlink system is a MIMO-OFDMA based system. We consider a MIMO–OFDM system with $M_T$ transmit and $M_R$ receive antennas.

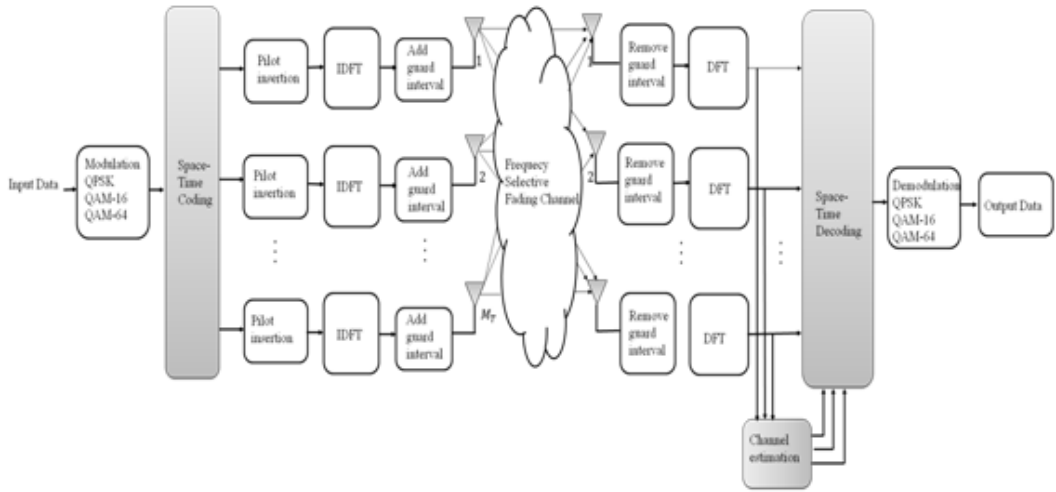

Figure 7: LTE MIMO-OFDM Model

OFDMA is defined as the multiple access technique for LTE Downlink systems. OFDMA is a multiple access scheme based on OFDM modulation technique. As shown in Fig. 3, we consider an OFDM system with $N_c$ carriers occupying the bandwidth. OFDM technique aims to split the data stream to be transmitted onto narrowband orthogonal subcarriers by the means of the Inverse Discrete Fourier Transform (IDFT) operation. In order to mitigate ISI and ICI, a guard interval is inserted with the length of either $L_{CP}$ or $L_{ZP}$ for respectively prefix cycling or zero padding is inserted at the beginning of each transmitted OFDM symbol. The inserted guard interval must be chosen to be equal or longer than the maximum excess delay of the channel in order to completely suppress ISI and ICI.

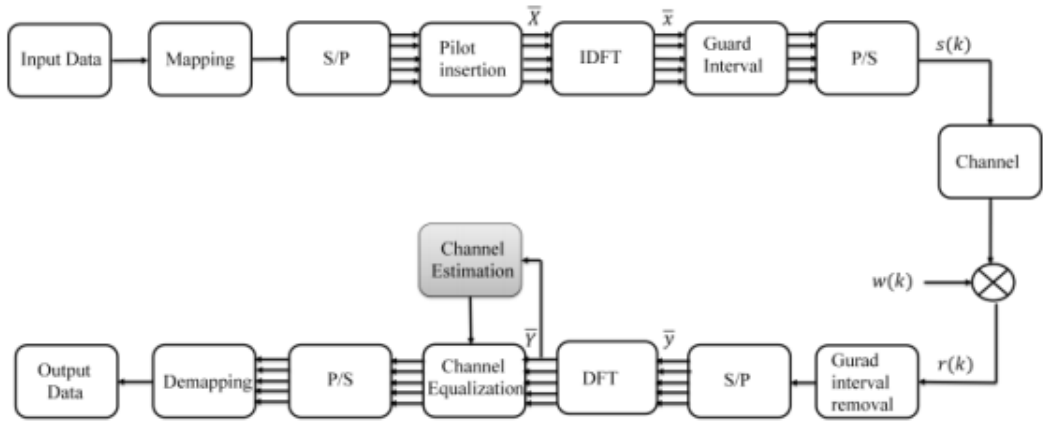

Figure 8: OFDM baseband system

OFDM symbols are sent over frequency selective multipath fading channels assumed independent of each other. The multipath fading channel can be modeled as a FIR (Finite Impulse Response) with *L* taps for each channel:

$$h(t,\tau) = \sum_{l=0}^{L-1} h_l(t)\, \delta(t - \tau_l) \tag{1}$$

$h_l$ and $\tau_l$ are respectively the impulse response and the multipath delays of the channel. It is sufficient to consider in our system model only a single transmit and a single receive.
At one receive antenna, and after removing the guard interval and performing the DFT, the received OFDM symbol can be written as:

$$\overline{Y} = DFT_N\big(IDFT_N(\overline{X}) \otimes \overline{h} + \overline{W}\big) = \underline{X}\,\overline{H} + \overline{W} \tag{2}$$

$\overline{Y}$ represents the received signal vector; $\underline{X}$ is a diagonal matrix. $\overline{H}$ is a channel frequency response vector and $\overline{W}$ is an additive complex-valued white Gaussian noise vector with zero mean and variance $\sigma_W^2$. The noise is assumed to be independent of the transmitted signals.

## 4. Channel estimation

In LTE Downlink systems, channel estimation is performed based on pilot signals called RSs (Reference Signals) [3]. Depending on the employed guard interval, RSs are transmitted during in the first OFDM symbol and the fifth OFDM symbol of every time slot for short guard interval while they are transmitted in the first and the fourth OFDM of every time slot for long guard interval. Figure 9 shows reference signal pattern for one antenna.

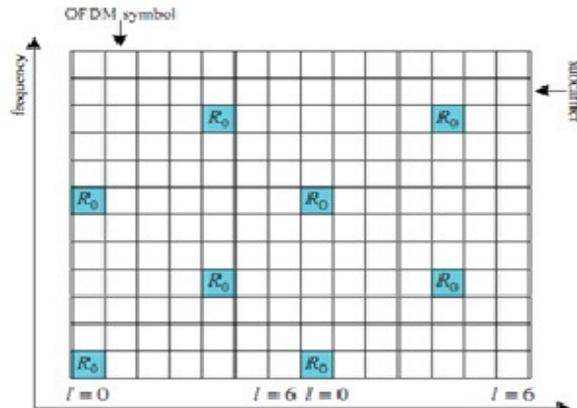

Figure 9: Reference signal pattern for one antenna [5]

From (2), the received pilot signals can be written as:

$$\overline{Y}_P = \underline{X}_P \overline{H}_P + \overline{W}_P \quad (3)$$

$(.)_P$ denotes positions where reference signals are transmitted.

### 4.1   Least Square (LS)

The least square estimates (LS) of the channel at the pilot subcarriers given in (3) can be obtained by the following equation [6]:

$$\overline{H}_P^{LS} = (\underline{X}_P)^{-1} Y_P \quad (4)$$

$\overline{H}_P^{LS}$ represents the least-squares (LS) estimate obtained over the pilot subcarriers.
 Without using any knowledge of the statistics of the channels, the LS estimators are calculated with very low complexity, but they suffer from a high mean-square error.

### 4.2   Linear Mean Minimum Square Error LMMSE

The LMMSE channel estimator is designed to minimize the estimation MSE .The LMMSE estimate of the channel responses given in (3) is [7]:

$$\overline{H}_P^{LMMSE} = R_{\overline{H}_P \overline{H}_P} \left( R_{\overline{H}_P \overline{H}_P} + \sigma_W^2 \left( \underline{X}_P \underline{X}_P^H \right)^{-1} \right)^{-1} \overline{H}_P^{LS} \quad (5)$$

$R_{\overline{H}_P \overline{H}_P}$ represents the autocorrelation matrix of the subcarriers with reference signals. The high complexity of LMMSE estimator (4) is due to the inversion matrix lemma .Every time data changes, inversion is needed. The complexity of this estimator can be reduced by averaging the transmitted data.  Therefore, we replace the term in (4) with its expectation.

The simplified LMMSE estimator becomes [7]:

$$\overline{H}_P^{LMMSE} = R_{\overline{H_P H_P}} \left( R_{\overline{H_P H_P}} + \frac{\beta}{SNR} I_P \right)^{-1} \overline{H}_P^{LS} \qquad (6)$$

where $\beta$ is scaling factor depending on the signal constellation (i.e for QPSK and for 16-QAM). SNR is the average signal-to-noise ratio, and $I_P$ is the identity matrix.

### 4.3 Low rank approximation

The advantage of the LMMSE estimator is the avoidance of the matrix inversion but at the high computational complexity. In order to improve the performance of channel estimation, the LMMSE estimator can be approximated by using a low-rank simplified LMMSE estimator achieved by the singular value decomposition.

The channel impulse response autocorrelation function matrix $R_{HpHp}$, which is the channel impulse response autocorrelation matrix which can be singular decomposed as [7]:

$$R_{\overline{H_P H_P}} = U \Lambda U^H \qquad (7)$$

where $U$ is the unitary matrix which consist of eigenvector, $\Lambda = \text{diag}(\lambda_1, \lambda_2, \ldots, \lambda_N)$ is a diagonal matrix which contains the eigenvalues of $R_{H_P H_P}$. The eigenvalues are ranged in descending order. The simplified LMMSE of order p can be written as:

$$\overline{H}_P^{Lr-LMME} = U \begin{bmatrix} \Delta_p & 0 \\ 0 & 0 \end{bmatrix} U^H \overline{H}_P^{LS} = U \Delta U^H \overline{H}_P^{LS} \qquad (8)$$

where $\Delta$ is a diagonal matrix defined as:

$$\Delta = \Lambda \left( \Lambda + \frac{\beta}{SNR} \right)^{-1}$$

$$= diag\left( \frac{\lambda_1}{\lambda_1 + \frac{\beta}{SNR}}, \frac{\lambda_2}{\lambda_2 + \frac{\beta}{SNR}}, \ldots, \frac{\lambda_N}{\lambda_N + \frac{\beta}{SNR}} \right) \qquad (9)$$

## 5. SIMULATION RESULTS

Simulation studies are conducted to investigate the performance of channel estimation techniques for CP-LTE and ZP-LTE Downlink systems. Simulations results are based on LTE-5 MHz Downlink system with 2 transmit and 2 receive antennas. As cited in Table1, the number of used subcarriers is 300. The length of the guard interval is set to 16. The transmitted signals are quadrature phase-shift keying (QPSK) modulated.100 radio frames are sent through a frequency-selective channel. The frequency-selective fading channel responses are randomly generated with a Rayleigh probability distribution. Table 2 gives a summary of simulation parameters.

Table 2: Simulation Parameters

| Simulation Parameters | |
|---|---|
| LTE Bandwidth | 5 MHz |
| Number of used subcarriers | 300 |
| Cyclic prefix length | 16 |
| Number of transmitted Frames | 100 |
| Number of transmit antenna | 2 |
| Number of Receive antenna | 2 |
| Modulation scheme | QPSK |
| Channel Model | Rayleigh |

As shown in Figure 10, CP-LTE systems outperform ZP-LTE systems in terms of BER. CP-LMMSE channel estimator looks the best channel estimator technique.

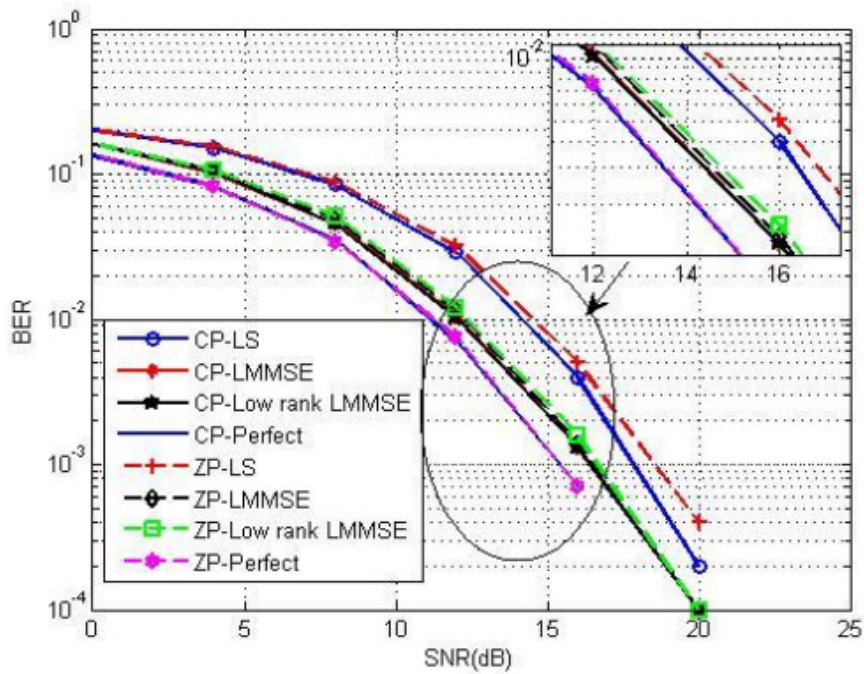

Figure 10: BER versus SNR

Figure 11 shows the performance of channel estimators for both CP-LTE and ZP-LTE systems. In terms of MSE, it is clear that each estimator in CP-LTE systems and its homologue in ZP-LTE systems have the same performance for low SNR values. However, channel estimation techniques for CP-LTE systems show better performance for high SNR values.

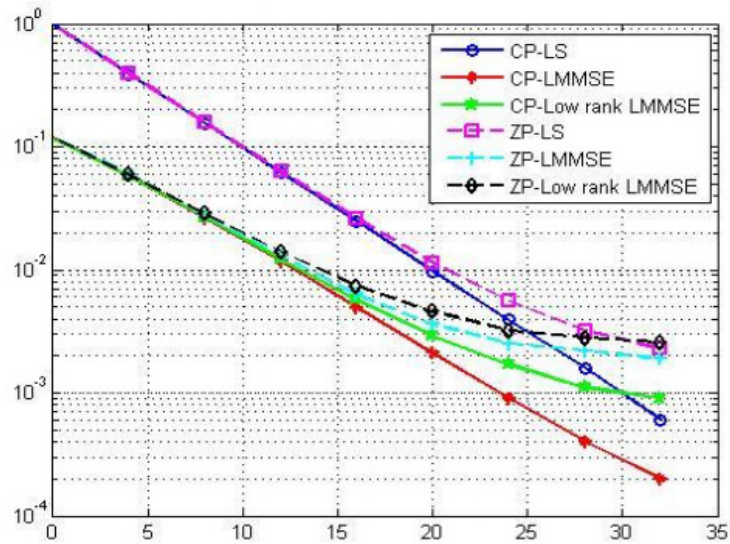

Figure 11: MSE versus SNR

## 6. Conclusion

In this paper, we have studied the performance of channel estimation techniques for two types of LTE systems: CP-LTE and ZP-LTE Downlink systems. Channel estimation is critical to receiver performance in LTE systems. LTE Downlink is based MIMO-OFDMA system. Most of the research works for MIMO-OFDM systems are based on the assumption that the inserted guard interval is based on CP technique. In order to avoid the power efficiency, we proposed to study the performance of LTE systems based on ZP technique. Simulation results have shown that CP-LTE systems outperform ZP-LTE in terms of BER. In terms of MSE, channel estimation techniques in the two systems show same performances but only for low SNR values. For high SNR values, channel estimators are better for CP-LTE Downlink systems.

**Authors**

**Abdelhakim KHLIFI** was born in Düsseldorf, Germany, on January 07, 1984. He graduated in Telecommunications Engineering, from National Engineering School of Gabès in Tunisia, July 2007. In June 2009, he received the master's degree of research in communication systems of the School of Engineering of Tunis ENIT. Currently he is a Ph.D student at the National School of Engineering of Tunis. His research spans radio channel estimation in LTE MIMO-OFDM systems.

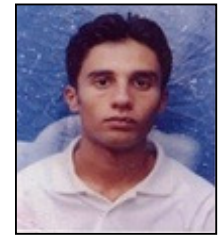

**Ridha BOUALLEGUE** is Professor at the National Engineering School of Tunis, Tunisia (ENIT). He practices at the Superior School of communication of Tunis (Sup'Com). He is founding in 2005 and Director of the Research Unit "Telecommunications Systems: 6'Tel@Sup'Com˅T. He is founding in 2005, and Director of the National Engineering School of Sousse. He received his PhD in 1998 then HDR in 2003. His research and fundamental development, focus on the physical layer of telecommunication systems in particular on digital communications systems, MIMO, OFDM, CDMA, UWB, WiMAX, LTE, has published 2 book chapters, 75 articles in refereed conference lectures and 15 journal articles (2009).

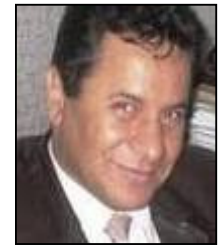